\documentclass[reprint,superscriptaddress,showpacs,nofootinbib,amsmath,amssymb,aps,prl]{revtex4-1}
\usepackage{graphicx}
\usepackage{dcolumn}
\usepackage{bm}
\usepackage[utf8]{inputenc}
\usepackage[frenchb]{babel}
\usepackage{braket}
\usepackage[T1]{fontenc}

\begin{document}

\preprint{APS/123-QED}

\title{Pairing of electro-magnetic bosons under spin-orbit coupling}

\author{S. V. Andreev}
\email[Electronic adress: ]{Serguey.Andreev@gmail.com}
\affiliation{National Research Center "Kurchatov Institute" B.P.\ Konstantinov Petersburg Nuclear Physics Institute, Gatchina 188300, Russia}

\date{\today}

\begin{abstract}

We discuss pairing of light-matter bosons under effective spin-orbit (SO) coupling in two-dimensional semiconductors. The SO coupling is shown to induce dynamical broadening of a two-body bound state. Application of a transverse magnetic field yields the rich Feshbach resonance phenomenology. We predict quantum bosonic halos with a synthetic angular momentum $L_z=\pm 2\hbar$. The d-wave-like dressing of the nominally s-wave bound state is induced by SO coupling to the continuum in the open channel. The fundamental properties of the emergent quantum number remain to be explored.           

\end{abstract}


\maketitle

Strongly correlated pairs of photons have been of paramount interest for fundamental and applied quantum science \cite{Kwiat1995, Boto2000, Lamas2001, Braunstein2005}. In the semiconductor technology pairing of electro-magnetic waves is enabled by formation of a biexcitonic molecule. Quantum entanglement of photon pairs has been experimentally demonstrated with biexcitons in semiconductor quantum dots \cite{Salter2010,Dousse2010}. Further progress toward greater functionalities would require controllable pairing in macroscopic ensembles of light-matter bosons \cite{Boto2000, Lamas2001, Braunstein2005}. Observation of biexcitons in several prominent classes of two-dimensional (2D) semiconductors \cite{XXinTMD1,Hao2017,XXinTMD3} has spurred an intense theoretical research in the field over the past years (see \cite{Andreev2020} and references therein).

Bosonic modes in planar semiconductors have fine structure due the long-range exchange interaction of the constituent fermions \cite{Sham, Gupalov, Glazov, Louie2015, Wang2015} and TE-TM splitting of the associated electro-magnetic field \cite{kaliteevskii, Kavokin2005, Bleu2017}. Both contributions can be regarded as interaction of the boson pseudospin with a momentum-dependent magnetic field. The influence of such effective spin-orbit (SO) coupling on the boson pairing is a challenging question, which has been addressed only scarcely in our recent studies \cite{Andreev2021}.

In this paper we present a powerful theoretical approach to the problem based on the method of second quantization. Assuming asymptotic completeness of the boson-boson scattering we construct a generic pairing Hamiltonian amenable to simple mean-field models for macroscopic ensembles and multichannel scattering formalism for two particles in vacuum. The SO interaction couples bound states to continua in the channels with different spin configurations. Application of a static magnetic field along $z$ (the Faraday geometry) yields a rich phenomenology akin to the Feshbach resonance in ultra-cold atomic gases \cite{Moerdijk1995}. We obtain analytical estimates of the positions of resonances for excitons and polaritons in atomically thin layers of transition metal dichalcogenides (TMD's). Close to the resonance we predict macroscopic bosonic molecules (the so-called halo states \cite{Jensen2004}) possessing an angular momentum $L_z=\pm 2\hbar$. The nominally s-wave bound state acquires a d-wave-like halo due to the SO coupling with the continuum. Straightforward extension of our formalism indicates that one may expect existence of analogous mechanisms for fermions. Fundamental properties of the emergent quantum number and its possible use in quantum technologies remain to be explored.

The electro-magnetic boson fields in a 2D semiconductor are guided (surface) exciton-polaritons propagating in the structure plane. At low momenta inside the light cone there is either rapid radiative decay \cite{Hanamura1988} or hybridization with a cavity mode \cite{Weishbuch1992}. The latter results in a strong reduction of the polariton effective mass. We shall neglect leakage of the polariton modes from the cavity. For excitons, we assume that their momenta are outside of the light cone.

The optical selection rules lock the photon spin to the spin orientation of an electron and a hole constituting the exciton. This allows one to treat interaction of the boson with an external magnetic field and different contributions to the effective SO fields on equal footing. The generic single-particle Hamiltonian reads
\begin{equation}
\label{SP}
\hat H_i=K(\hat {\bm p_i}) \hat{\mathrm 1}_i-\hbar \bm\omega_i\cdot \hat{\bm S}_i
\end{equation}   
with
\begin{equation}
\bm\omega=\bm\Omega(\hat {\bm p})+\frac{\mu_B}{\hbar}gB\bm n_z
\end{equation}
being a superposition of the external field and the in-plane SO field 
\begin{equation}
\label{SO}
\bm\Omega(\hat {\bm p})=\Omega_x(\hat {\bm p}) \bm n_x+\Omega_y(\hat {\bm p}) \bm n_y
\end{equation}
constructed on the basis of the momentum operator $\hat {\bm p}=(\hat p_x,\hat p_y)$. The index $i$ in \eqref{SP} is introduced to label the particles. Bearing in mind TMD's as primary experimental testbeds, we keep only the $z$-component of the external magnetic field. The exciton $g$-factor in such materials is due to out-of-plane orbital magnetic moments of the valence and conduction bands \cite{Sallen2012, Zeng2012, Srivastava2015}. For polaritons at the bottom of their dispersion $g$ is small and variation of their energy with $B$ is due to the exciton diamagnetic shift \cite{Deveaud2015, Crooker2018}, which may be conveniently absorbed into the kinetic energy operator $K(\hat {\bm p})$. The latter also includes the momentum-dependent shifts ensuring monotonous growth of the genuine single-particle dispersions. The photon spin
\begin{equation}
\label{spin}
\hat{\bm S}=\hat\sigma_x \bm n_x +\hat\sigma_y \bm n_y+\hat\sigma_z \bm n_z
\end{equation}
is the spin-1 operator of a transverse electromagnetic field. Here $\hat\sigma_x$, $\hat\sigma_y$ and $\hat\sigma_z$ are Pauli matrices, and $\bm n_x$, $\bm n_y$, $\bm n_z$  form an orthonormal basis. We shall adopt the notation $\ket{\uparrow}$ and $\ket{\downarrow}$ for the basis states characterized by $S_z=+1$ and $S_z=-1$, respectively. The three components of an expectation value $\braket{\hat{\bm S}}$ are the Stokes parameters encoding the photon polarization. It follows that under time reversal $\hat S_{x,y}\rightarrow \hat S_{x,y}$ \cite{footnote1} and, therefore,
\begin{equation}
\label{TR}
\bm\Omega(\hat {\bm p})=\bm\Omega(-\hat {\bm p}),
\end{equation}
so that the Hamiltonian \eqref{SP} remains intact. Explicitly, we shall assume
\begin{subequations}
\label{EffectiveExcitonFields}
\begin{align}
K(\bm p)&=\frac{\hbar^2p^2}{2m}+\frac{\hbar\upsilon p}{2}\\
\bm\Omega(\bm p)&=-\frac{\upsilon p}{2}(\bm n_x\cos{2\theta}+\bm n_y\sin{2\theta})
\end{align}
\end{subequations}
for excitons \cite{Sham, Gupalov, Glazov, Louie2015, Wang2015, Andreev2021} and
\begin{subequations}
\label{EffectivePolaritonFields}
\begin{align}
\label{EffectivePolaritonFields1}
K(\bm p)&=\frac{\hbar^2p^2}{2m_\ast}+\frac{\delta(B)}{2}\\
\bm\Omega(\bm p)&=-\frac{\hbar p^2}{2m_\mathrm{LT}}(\bm n_x\cos{2\theta}+\bm n_y\sin{2\theta})
\end{align}
\end{subequations}
for polaritons \cite{kaliteevskii, Kavokin2005, Bleu2017, Andreev2021}, where $m_\ast^{-1}=m_\mathrm{LP}^{-1}+m_\mathrm{LT}^{-1}$ with $m_\mathrm{LP}$ and $m_\mathrm{LT}$ being the lower polariton mass and the parameter characterizing the longitudinal-transverse splitting of the polariton dispersion, respectively. The shift $\delta(B)/2$ includes the $B=0$ detuning between the cavity and the exciton modes and the exciton diamagnetic shift \cite{Deveaud2015}.

Consider now a system of two particles. There are four basis states $\ket{\uparrow\uparrow}$, $\ket{\uparrow\downarrow}$, $\ket{\downarrow\uparrow}$, $\ket{\downarrow\downarrow}$ whose linear combinations realize the states $S_z=+2,0,-2$ of the total spin $S_z=S_{z,1}+S_{z,2}$. We notice that since $\bm\Omega(\hat {\bm p}_i)$ lies in the structure plane, the sum $\bm\Omega(\hat {\bm p}_1)\cdot \hat{\bm S}_1+\bm\Omega(\hat {\bm p}_2)\cdot \hat{\bm S}_2$ does not commute with $S_z^2$. Hence, the SO coupling may change the spin state of a pair by flipping the spin of either of the two particles. This property in combination with the symmetry relation \eqref{TR} are the key ingredients for the phenomenology presented in this work.

To proceed, let us introduce the (free) pair creation and annihilation operators
\begin{equation}
\label{FreePairs}
\begin{split}
\hat C^\dagger_{\sigma_1\sigma_2,\bm k,\bm K}&=\hat a^\dagger_{\sigma_1 \bm k+\bm K/2} \hat a^\dagger_{\sigma_2, -\bm k+\bm K/2}\\
\hat C_{\sigma_1\sigma_2,\bm k,\bm K}&=\hat a_{\sigma_2, -\bm k+\bm K/2}\hat a_{\sigma_1 \bm k+\bm K/2},
\end{split}
\end{equation}
where $\bm k=(\bm p_1-\bm p_2)/2$ and $\bm K=\bm p_1+\bm p_2$ are the relative and center-of-mass (c.o.m.) momenta, respectively, and $\sigma=\uparrow,\downarrow$ labels the particle spin. The particle creation (annihilation) operators $\hat a_{\sigma, \bm p}^\dagger$ ($\hat a_{\sigma, \bm p}$) obey the usual boson commutation relations. It follows that
\begin{equation}
\hat C_{\sigma_1\sigma_2 ,\bm k,\bm K}=\hat C_{\sigma_2\sigma_1 ,-\bm k,\bm K}.
\end{equation}
Next, we define the molecular operators
\begin{equation}
\label{BoundStates}
\hat C_{\uparrow\downarrow, n, \bm K}=\sum_{\bm k} \phi_n(\bm k)\hat C_{\uparrow\downarrow, \bm k,\bm K},
\end{equation}           
where the function $\phi_n(\bm k)$ is the Fourier image of the molecular wavefunction of the relative motion
\begin{equation}
\label{phi}
\varphi_n(\bm r)=\frac{1}{\sqrt{S}}\sum_{\bm k}\phi_n(\bm k) e^{i\bm k\bm r},
\end{equation}
with the index $n$ standing for a full set of possible quantum numbers and $S$ being the quantization area. Binding of light-matter bosons occurs via their excitonic component due to exchange of the identical fermions. A close analogue of the biexciton is the positronium molecule $\mathrm{Ps}_2$ \cite{Cassidy2018}. In the spirit of covalent bonding in the molecular hydrogen $\mathrm{H}_2$, the singlet electrons produce Coulomb attraction for holes and vice versa. Opposite orientations of the fermionic spins imply the singlet configuration for the associated photons. Only even spin and orbital boson wave functions are allowed \cite{Landau1965, Yang1950}. The wave function $\varphi_n(\bm r)$ can be obtained as a solution of the Schrodinger equation for the relative motion
\begin{equation}
\label{BoundStateEq}
[K_\mathrm{rel}(\hat{\bm k})+V_{\uparrow\downarrow}(r)]\varphi_n(\bm r)=\varepsilon_n \varphi_n(\bm r),
\end{equation}
where $V_{\uparrow\downarrow}(r)$ is a potential describing interaction of two composite bosons \cite{Ivanov1998}. The kinetic energy of relative motion $K_\mathrm{rel}(\hat{\bm k})$ is defined in a standard way via the rearrangement $K(\bm {\hat p}_1)+K(\bm {\hat p}_2)=K_\mathrm{rel}(\hat{\bm k})+K_\mathrm{cm}(\hat{\bm K})$ which always holds at least in the c.o.m. reference frame.

In terms of the free pair and molecular operators the second-quantized \textit{pairing Hamiltonian} can be written as
\begin{widetext}
\begin{equation}
\label{PairingHamiltonian}
\hat H=\frac{1}{2}\sum\limits_{\sigma_1^\prime\sigma_2^\prime\sigma_1\sigma_2}\sum\limits_{\bm k^\prime,n^\prime,\bm k,n}\sum_{\bm K^\prime,\bm K}\mathcal H^{\bm k^\prime (n^\prime),\bm K^\prime,\bm k (n), \bm K}_{\sigma_1^\prime\sigma_2^\prime\sigma_1\sigma_2} \hat C^\dagger_{\sigma_1^\prime\sigma_2^\prime,\bm k^\prime (n^\prime),\bm K^\prime}\hat C_{\sigma_1\sigma_2 ,\bm k (n),\bm K},
\end{equation}
\end{widetext}
where we have used the obvious notation $\hat C_{\sigma_1\sigma_2 ,n,\bm K}\equiv\delta_{\sigma_1\bar\sigma_2}\hat C_{\uparrow\downarrow, n, \bm K}$. From a mathematical viewpoint, the form \eqref{PairingHamiltonian} is dictated by the principle of asymptotic completeness of the resultant scattering theory \cite{Taylor}. This fundamental principle is expected to hold in our case on timescales shorter than the boson radiative lifetime. The matrix elements
\begin{widetext}
\begin{equation}
\mathcal H^{\bm k^\prime (n^\prime),\bm K^\prime,\bm k (n), \bm K}_{\sigma_1^\prime\sigma_2^\prime\sigma_1\sigma_2}=\int\psi^*_{\bm k^\prime (n^\prime),\bm K^\prime}(\bm r,\bm R)\bra{\sigma_1^\prime\sigma_2^\prime}\mathcal {\hat H}\ket{\sigma_1\sigma_2} \psi_{\bm k (n),\bm K}(\bm r,\bm R)d\bm r d\bm R
\end{equation}
\end{widetext}
are taken on the free and bound state wave functions
\begin{subequations}
\begin{align}
\psi_{\bm k\bm K}(\bm r,\bm R)&=\frac{1}{S}e^{i\bm k\bm r+i\bm K\bm R}\\
\psi_{n,\bm K}(\bm r,\bm R)&=\frac{1}{\sqrt{S}}e^{i\bm K\bm R}\varphi_n(\bm r).
\end{align}
\end{subequations}
The $4\times4$ matrix Hamiltonian $\mathcal {\hat H}$ is given by
\begin{equation}
\mathcal{\hat H}=\hat H_1\oplus\hat H_2+\mathcal{\hat V},
\end{equation}
where "$\oplus$" denotes the Kronecker sum, the $2\times2$ matrices $\hat H_i$ are given by Eq. \eqref{SP} and
\begin{equation}
\mathcal{\hat V}=\sum_{\sigma,\sigma^\prime} V_{\sigma\sigma^\prime}(r)\ket{\sigma\sigma^\prime}\bra{\sigma\sigma^\prime}.
\end{equation}
Strictly speaking, the two-body interaction potentials $V_{\sigma\sigma^\prime}(r)$ are well-defined only at inter-particle distances $r$ much larger than the exciton Bohr radius $a_\mathrm{X}$. This is sufficient, however, for considerations involving the low-energy scattering assumed in our study. Moreover, unless we are interested in long-range physics due to a possible permanent exciton dipole moment \cite{Andreev2016}, the actual shape of the potential curves is not important even at $r\gg a_\mathrm{X}$. We shall only assume that the potential $V_{\uparrow\downarrow}(r)$ possesses a bound state [Eq. \eqref{BoundStateEq}], whereas $V_{\uparrow\uparrow}(r)$ and $V_{\downarrow\downarrow}(r)$ do not.

In practice, excited states of a biexciton have never been observed. We are also unaware of any theoretical claim of such states. In fact, studies of the analogous dipositronium problem suggest that such states may actually not exist due to a very diffuse structure of a four-particle complex consisting of electrons and holes with equal masses \cite{Varandas2015}. We shall therefore assume that the set of bound states labelled by $n$ in Eq. \eqref{PairingHamiltonian} consists of a single $s$-wave (ground) state. This assumption, while not implying any loss of generality, will also highlight the emergent nature of the new state derived below.

Let us apply our pairing formalism to excitons. The external magnetic field splits the triplet of the scattering channels with $S_z=-2,0,+2$ by the amount $2\mu_B g B$. When $B$ crosses the threshold value $B_r$ (to be derived below), the lowest-energy scattering states ($S_z=+2$) come into resonance with the $S_z=0$ bound state. Assuming $\lvert B-B_r\rvert\ll B_r$ one may neglect the coupling to the $S_z=0,-2$ scattering states (as well as to other possible states such as dark excitons \cite{Molas2019})  and reduce the generic Hamiltonian \eqref{PairingHamiltonian} to
\begin{equation}
\label{ExcitonHamiltonian}
\begin{split}
&\hat H_\mathrm{X}=\sum_{\bm k}[K(\bm k)-\mu_B g B]\hat a^\dagger_{\uparrow,\bm k}\hat a_{\uparrow,\bm k}+\varepsilon \hat C_{\uparrow\downarrow}^\dagger  \hat C_{\uparrow\downarrow}\\
&+\frac{1}{2S}\sum_{\bm k,\bm k^\prime}\hat a_{\uparrow,\bm k^\prime}^\dagger \hat a_{\uparrow,-\bm k^\prime}^\dagger V_{\uparrow\uparrow}(\bm k^\prime-\bm k)\hat a_{\uparrow,-\bm k} \hat a_{\uparrow,\bm k}\\
&-\sum_{\bm k} \hbar \bm\Omega^{(s)}(\bm k)\cdot [\bm S_{\downarrow\uparrow}\phi^\ast (\bm k)\hat C_{\uparrow\downarrow}^\dagger \hat a_{\uparrow,-\bm k} \hat a_{\uparrow,\bm k}+\mathrm{H.c.}],
\end{split}
\end{equation} 
where
\begin{equation}
\label{SymmOmega}
\bm\Omega^{(s)}(\bm k)=\frac{\bm\Omega(\bm k)+\bm\Omega(-\bm k)}{2},
\end{equation}            
$\bm S_{\downarrow\uparrow}\equiv \bra{\downarrow} \bm{\hat S}\ket{\uparrow}$ with $\bm{\hat S}$ defined by Eq. \eqref{spin},  $\hat C_{\uparrow\downarrow}$ is the shortcut for the molecular operator $\hat C_{\uparrow\downarrow, n=1s, \bm K=0}$ and, as usual, we have assumed that all pairs are at $\bm K=0$. One may readily recognize the structure of the Fano-Anderson model of a discrete level in a continuum \cite{Fano1961}. Mean-field solutions of analogous models have provided much insight into the collective behaviour of ultra-cold atoms with Feshbach resonances \cite{Timmermans1999, Weichman2004, Gurarie2007}. The important difference of our system from atomic settings and previously considered dipolar excitons featuring a shape resonance \cite{Andreev2016} is the dynamical (orbital) nature of the coherent coupling between the open ($S_z=+2$) and closed ($S_z=0$) channels. For two particles in vacuum the Hamiltonian \eqref{ExcitonHamiltonian} yields the on-shell $T$-matrix
\begin{equation*}
T_{\uparrow\uparrow}=T^{(\mathrm{bg})}_{\uparrow\uparrow}+T^{(\mathrm{res})}_{\uparrow\uparrow}
\end{equation*}
with $T^{(\mathrm{bg})}_{\uparrow\uparrow}$ being the standard background contribution due to the potential $V_{\uparrow\uparrow}(\bm k^\prime-\bm k)$ and
\begin{equation}
T^{(\mathrm{res})}_{\uparrow\uparrow}(\bm k^\prime,\bm k, E_{\bm k}+i0)=\frac{2\lvert \tilde\phi(\bm k)\rvert^2\lvert \hbar \bm\Omega^{(s)}(\bm k)\cdot \bm S_{\uparrow\downarrow}\rvert^2 e^{2i(\theta^\prime-\theta)}}{E_{\bm k}-\Delta-\Pi(E_{\bm k}+i0)},
\end{equation}
where $\Delta\equiv\varepsilon+2\mu_B gB$ is the magnetic-field-induced detuning between the channels (note that $\varepsilon<0$), $E_{\bm k}\equiv K(\bm k)+K(-\bm k)$ is the kinetic energy of the relative motion of two excitons, $\tilde\phi(\bm k)\equiv \sqrt{S} \phi(\bm k)/2\pi$ with $\phi(\bm k)$ defined by Eq. \eqref{phi} and the pair-bubble reads 
\begin{equation}
\Pi(E_{\bm k}+i0)=\int \frac{2\lvert \tilde\phi(\bm q)\rvert^2\lvert \hbar \bm\Omega^{(s)}(\bm q)\cdot \bm S_{\uparrow\downarrow}\rvert^2}{E_{\bm k}-E_{\bm q}+i0}d\bm q.
\end{equation}
The $d$-wave-like dependence of $T^{(\mathrm{res})}_{\uparrow\uparrow}$ on the scattering angle is entirely due to the SO coupling. The bare excitonic molecule has zero orbital momentum. By using the explicit formulas \eqref{EffectiveExcitonFields} and assuming $k\ll m\upsilon$ we find for the pole of the $T$-matrix
\begin{equation}
\label{ExcitonResonance}
E=\Delta-\frac{\pi}{3}E_a-\frac{\pi}{2}E-\pi \frac{E^2}{E_a}-\beta_E \ln\left(\frac{E_a-E}{E}\right)-i\pi\beta_E,
\end{equation} 
where $\beta_E=\pi E^3/E_a^2$ and $E_a\equiv \hbar \upsilon/a$ with $a$ being the molecular radius (on the order of $a_\mathrm{X}$). We have used the standard \textit{ansatz} $\lvert \tilde\phi(\bm k)\rvert= a \mathrm{\theta} (a^{-1}-k)$ with $\mathrm{\theta}(x)$ being the Heaviside theta function. Solution of Eq. \eqref{ExcitonResonance} at $E\rightarrow 0$ can be recast in the form
\begin{equation}
\label{ExcitonResonanceB}
E(B)=\frac{4}{3\pi}\mu_B g(B-B_r),
\end{equation}
where
\begin{equation}
B_r= (\lvert \varepsilon\rvert+\pi/3 E_a)/2\mu_B g.
\end{equation}
The solution \eqref{ExcitonResonanceB} is a resonance at $B>B_r$ and a \textit{synthetic} bound state of the two-channel model \eqref{ExcitonHamiltonian} at $B<B_r$. Although the density of states in 2D remains finite at zero energy, the resonance has vanishing width $\beta_E$ due to the orbital origin of the effective magnetic field switching the pair spin configuration. The slope of the line $E(B)$ defines the relative weight of the bare molecule in the (normalized) wave function of the synthetic bound state
\begin{equation}
\label{PsiX}
\ket{\Psi(\bm r)}=\Upsilon (\bm r)\ket{\uparrow\uparrow}+w\varphi (\bm r)\tfrac{1}{2}(\ket{\uparrow\downarrow}+\ket{\downarrow\uparrow}) 
\end{equation}
via the identity $w^2=(2\mu_B g)^{-1}\partial E(B)/\partial B$. In the vicinity of $B_c$, where Eq. \eqref{ExcitonResonanceB} holds, we obtain that only around $20$ $\%$ of the total probability density remains in the bare $s$-wave biexciton. The dominant contribution to the wave function \eqref{PsiX} is due to the quantum halo
\begin{equation}
\Upsilon (\bm r)=\frac{w}{\sqrt{2}\pi}\int \hbar \bm\Omega^{(s)}(\bm q)\cdot \bm S_{\uparrow\downarrow} G_{\uparrow\uparrow}(0)\tilde\phi(\bm k)e^{-i\bm q\bm r}d\bm q,
\end{equation}
where $G_{\uparrow\uparrow}(E)=(E-E_{\bm q})^{-1}$ is the Green function of the open channel. At $r\gg a$ we obtain
\begin{equation}
\label{halo}      
\Upsilon (\bm r)\propto a^{-1}\left(\frac{a}{r}\right)^{3/2} \cos \left(\frac{r}{a}\right)e^{-2i\alpha},
\end{equation}
where $\alpha$ is the polar angle of $\bm r$. In contrast to the strongly localized $s$-wave core, the halo decays algebraically and exhibits density oscillations in the radial direction. Most remarkably, the halo carries an angular momentum $L_z=-2\hbar$ (opposite to the applied magnetic field). We suggest that the emergent quantum number is due to conversion of the core spin fluctuations into the peripheric orbital motion by the SO coupling. Possibly the same mechanism underlies the formation of polarized stripes in a resonantly paired superfluid of dipolar excitons \cite{Andreev2021}.

The energy $E(B)$ defined by Eq. \eqref{ExcitonResonanceB} should not be confused with the dissociation energy of the new state. The synthetic wave function \eqref{PsiX} itself represents a partially disintegrated state, the halo being a quantum superposition of the continuum states in the open channel. Destruction of the halo occurs when the thermal energy $k_B T$ becomes comparable with the Josephson energy of the coherent SO link between the channels, the latter being on the order of $E_a=\hbar\upsilon/a$ . For excitons in a TMD monolayer we estimate $E_a/k_B\sim 150$ K. The halo should emit entangled pairs of photons carrying an orbital angular momentum $\pm2\hbar$ due to leakage of the low-$k$ excitons into the light cone. Such photons may be detected by using the coincidence circuit supplemented with the fork-like interference holograms \cite{Mair2001}. We expect $B_r$ on the order of few tens of teslas which is within reach of the existing experimental facilities \cite{Crooker2018}.

For polaritons [Eqs. \eqref{EffectivePolaritonFields}] we assume that the detuning $\lvert\delta(B)\rvert\sim\lvert\varepsilon\rvert$ and greatly exceeds the vacuum Rabi splitting $\hbar \Omega_R$ \cite{Deveaud2015}. In this case one may apply the so-called giant oscillator strength model \cite{Ivanov1998} to treat the bipolariton as a biexciton for $\lvert \delta(B)-\varepsilon\rvert\gg \hbar \Omega_R$ \cite{Carusotto2010}. In the narrow range $\lvert \delta(B)-\varepsilon\rvert\lesssim \hbar \Omega_R$ the bipolariton is a loosely bound pair of cavity photons \cite{Rocca}. We exclude that latter regime from our present consideration. By substituting Eq. \eqref{SP} with $g\equiv0$ into the pairing Hamiltonian \eqref{PairingHamiltonian} we obtain
\begin{equation}
\label{PolaritonHamiltonian}
\begin{split}
&\hat H_\mathrm{P}=\sum_{\bm k,\sigma,\sigma^\prime}[K(\bm k)\delta_{\sigma\sigma^\prime}-\hbar\bm\Omega(\bm k)\cdot \bm S_{\sigma\sigma^\prime}]\hat a^\dagger_{\sigma,\bm k}\hat a_{\sigma^\prime,\bm k}+\varepsilon \hat C_{\uparrow\downarrow}^\dagger  \hat C_{\uparrow\downarrow}\\
&+\frac{1}{2S}\sum_{\bm k,\bm k^\prime,\sigma,\sigma^\prime}\hat a_{\sigma^\prime,\bm k^\prime}^\dagger \hat a_{\sigma,-\bm k^\prime}^\dagger V_{\sigma\sigma^\prime}(\bm k^\prime-\bm k)\hat a_{\sigma,-\bm k} \hat a_{\sigma^\prime,\bm k}\\
&-\sum_{\bm k} \hbar \bm\Omega^{(s)}(\bm k)\cdot [\bm S_{\downarrow\uparrow} \hat a_{\downarrow,\bm k}^\dagger \hat a_{\downarrow,-\bm k}^\dagger+\bm S_{\uparrow\downarrow} \hat a_{\uparrow,\bm k}^\dagger \hat a_{\uparrow,-\bm k}^\dagger]\phi(\bm k)\hat C_{\uparrow\downarrow}+\mathrm{H.c.},
\end{split}
\end{equation}                    
where again $\bm\Omega^{(s)}(\bm k)$ is given by Eq. \eqref{SymmOmega}. The massive character of the polariton kinetic energy [Eq. \eqref{EffectivePolaritonFields1}] allows us to define the scattering amplitude $f_{\sigma\sigma^\prime,\bm k}(\bm k^\prime)=-(2\pi)^2m_{\ast}/2\hbar^2 T_{\sigma\sigma^\prime}(\bm k^\prime,\bm k, E_{\bm k}+i0)$. For the resonant contribution we obtain
\begin{equation}
f^{(\mathrm{res})}_{\uparrow\uparrow,\bm k}(\bm k^\prime)=-\frac{2\pi e^{2i(\theta^\prime-\theta)}}{\tfrac{\tilde E_{\bm k}-\tilde\Delta}{\eta E_{\bm k}^2/E_a}+\ln (E_a/E_{\bm k}-1)+i\pi } 
\end{equation}
and $f^{(\mathrm{res})}_{\downarrow\downarrow,\bm k}(\bm k^\prime)$ can be obtained by the time reversal. The amplitude has the genuine $d$-wave form. Here $\eta\equiv \pi (m_\ast/m_\mathrm{LT})^2$, $E_a\equiv\hbar^2/m_\ast a^2$, $\tilde E_{\bm k}=(1+\eta) E_{\bm k}$ and $\tilde\Delta=\varepsilon-\delta (B)-\eta E_a/2$ (note that both $\varepsilon$ and $\delta (B)$ are negative). The position of the resonance is now given by
\begin{equation}
B_r=B_0+\eta E_a (2 \partial \lvert\delta(B)\rvert/\partial B)^{-1}
\end{equation}
with $B_0$ being defined by $\delta (B_0)=\varepsilon$. At $B\rightarrow B_r$ one has the synthetic state
\begin{equation}
\label{PsiP}
\ket{\Psi(\bm r)}=\Upsilon (\bm r)\ket{\uparrow\uparrow}+\Upsilon^\ast (\bm r)\ket{\downarrow\downarrow}+w\varphi (\bm r)\tfrac{1}{2}(\ket{\uparrow\downarrow}+\ket{\downarrow\uparrow}) 
\end{equation}
composed of a biexciton core with the relative weight $w^2=(1+\eta)^{-1}$ and a polariton halo. We have neglected weak SO coupling in the low-energy continuum. At $r\gg a$ the spatial component  $\Upsilon (\bm r)$ has the asymptotic form \eqref{halo}. The halo now has zero total angular momentum, but is entangled in both the polarization and angular momentum states of individual polaritons.

Finally, we point out possible existence of analogous mechanisms for fermions. Replacement of the commutation relations results in a sign change in Eq. \eqref{SymmOmega}. Consistently, the time reversal invariance now dictates that $\bm\Omega (\bm k)=-\bm\Omega (-\bm k)$, which is the case, in particular, for the Rashba \cite{Rashba1984} and Dresselhaus \cite{Dyakonov1986} SO couplings. The second-quantized forms of the type \eqref{ExcitonHamiltonian} and \eqref{PolaritonHamiltonian} then imply intriguing mean-field possibilities associated with topological properties of the Bogoliubov transformation \cite{Green2000}.

\bibliography{References}

\begin{thebibliography}{47}
\bibitem [{foo()}]{footnote1}%
  \BibitemOpen
  \href@noop {} {}\bibinfo {note} {Consider
  $S_x=\psi^\ast_\uparrow\psi_\downarrow+\psi^\ast_\downarrow\psi_\uparrow$.
  Under time reversal the components of a spinor transform as
  $\psi_{s,\sigma}\rightarrow\psi_{s,-\sigma}^\ast(-1)^{s+\sigma}$. In our case
  $s=1$ and $\sigma=+1,-1$. One obtains
  $\psi_\uparrow\rightarrow\psi_\downarrow^\ast$ and
  $\psi_\downarrow\rightarrow\psi_\uparrow^\ast$.}\BibitemShut {Stop}%
\end{thebibliography}
\end{document}